\documentstyle[seceq,supplement]{ptptex}

\input epsf.tex
\epsfverbosetrue

\newcommand{\cit}{\cite}

\newcommand{\bib}{\bibitem}

\newcommand{\nc}{\newcommand}
\nc{\id}{{\bf 1}}

\newcommand{\be}{\begin{equation}}
\newcommand{\ee}{\end{equation}}
\newcommand{\bea}{\begin{eqnarray}}
\newcommand{\eea}{\end{eqnarray}}
\newcommand{\ra}{\rightarrow}

\nc{\ie}{{\em i.e.}}
\nc{\eg}{{\em e.g.}}
\nc{\etal}{{\em et al.}}
\nc{\calF}{{\cal F}}
\nc{\calL}{{\cal L}}

\newcommand{\calS}{{\cal S}}

\renewcommand{\today}{January 31, 1998}
\nc{\Tr}{{\rm Tr\,}}
\nc{\half}{{\textstyle \frac{1}{2}}}

\title{
Abelian Projection without Ambiguities%
\protect\footnote{ETH preprint SCSC-TR-98-01.
Invited talk at YKIS\,'97 (Kyoto, Japan, December 1997).}
}

\author{
A.J. {\sc van der Sijs}\protect\footnote{E-mail address: arjan@scsc.ethz.ch}
}

\inst{
Swiss Center for Scientific Computing, ETH-Z\"urich\\
ETH-Zentrum, CH-8092 Z\"urich, Switzerland
}

\recdate{
\today
}

\abst{
Laplacian Abelian Projection is discussed.
This term refers to the use of a new (``Laplacian'') gauge fixing
prescription for implementing the Abelian Projection of QCD.
The gauge condition is based on the lowest-lying eigenvector of the
covariant Laplacian operator in the adjoint representation.

This Laplacian gauge fixing procedure is free of the ambiguities which plague
lattice simulations which work with the popular Maximally Abelian Gauge.
Furthermore, Laplacian gauge fixed configurations enjoy
a natural kind of smoothness.
These two properties are crucial for a reliable determination of physical
quantities using the Abelian Projection.

We also examine a new, Higgs-field-like observable which
emerges as a by-product of the method.
This quantity can be used to identify magnetic monopoles in a way
independent of the traditional prescription.
It is argued that physically relevant magnetic monopoles are
accomodated well by the Laplacian method, while they are suppressed
(too) strongly in Maximally Abelian Gauge.

Finally, first evidence of abelian dominance in the Laplacian Abelian
Projection is presented.
}

\begin{document}

\maketitle

\section{Panorama}

\label{panorama}

't Hooft's Abelian Projection paper \cit{tHo81} has been the point of
departure for many simulations of lattice gauge theories in the past decade.
His idea was that (partial) gauge fixing could expose the degrees of
freedom relevant in the long-distance regime of QCD.
He showed that if an abelian subgroup U(1)$^{N-1}$ of the non-abelian
gauge group SU(N) is left unfixed,
one obtains an abelian gauge theory whose gauge field content
consists of ``electrically'' charged vector fields and ``magnetic''
monopoles in addition to the $N-1$ abelian ``photons''.
The magnetic monopoles arise as defects in the gauge fixing.

An example of such a partial gauge condition in the continuum
is given by the covariant gauge\cit{tHo81}%
\protect\footnote{We shall restrict ourselves to
SU(2) throughout this presentation.}
\be
\sum_\mu \left( \partial_\mu \,\mp\, i A^3_\mu \right)
 \, A^\pm_\mu \ =\ 0 \, .
 \label{magcont}
\ee
In this gauge, the non-abelian components $A^{1,2}$ are covariantly
constant with respect to the abelian subgroup which is taken in the
3-direction.  There is no gauge condition on the abelian component $A^3$.

On the lattice, the most popular gauge is the ``Maximally Abelian Gauge''
(MAG)\cit{Kro87b},
which is defined as the configuration 
$\{ U_{\mu,x}^{(\Omega)} \}$ for which the gauge transformation
$\{ \Omega_x \}$ minimizes the functional
\be
\tilde\calS_U(\Omega) \ =\ \sum_{x,\mu} \left\{ 1 -
 \half \Tr \left[\sigma_3 U_{\mu,x}^{(\Omega)}
            \sigma_3 U_{\mu,x}^{(\Omega)\,+} \right] \right\} \, .
 \label{maglat}
\ee
The lattice gauge field $U_{\mu,x}$ transforms as
$U_{\mu,x}^{(\Omega)} \ =\ \Omega_x U_{\mu,x} {\Omega^+}_{\!\!\!\!\!x+\hat\mu}$.

These gauge conditions (\ref{magcont},\ref{maglat}) are in fact
related\cit{Smi91}.
In the naive continuum limit, Eq.\ (\ref{maglat}) becomes
\be
\tilde\calS_A(\Omega) \ = \ \frac12
  \int_V \left[ \left(A_\mu^{(\Omega)\,1}\right)^2
 \ + \ \left(A_\mu^{(\Omega)\,2}\right)^2 \right]
 \, ,
 \label{Scont}
\ee
and the stationarity condition
which its minimum has to satisfy is precisely Eq.\ (\ref{magcont}).

Many interesting results have been obtained using the MAG, as we will
briefly review in Sect.\ \ref{silver}.
This cannot be said of most other Abelian Projection gauges.
The reason is presumably that the MAG favours relatively
smooth gauge fields, unlike these other gauges.
There are a number of important drawbacks associated to the MAG, however.

Most importantly, it is impossible to implement the gauge condition
unambiguously, in practice.
This means, for example, that the result depends on the gauge in which
one happens to start the gauge fixing algorithm.
In other words, gauge invariance is lost.

Secondly, magnetic monopoles are not identified according to 't~Hooft's
prescription, by looking for singularities in the gauge fixing
(in fact one never attempts to find those singularities in the MAG!),
but instead using the U(1) procedure based on elementary cubes\cit{DeG80}.
It is not clear if there exists an exact relation between the monopoles
in the two definitions.

The Laplacian method discussed here manifestly solves the first problem.
In addition, it seems to induce a natural solution to the second
drawback: we find evidence that the gauge fixing singularities
are correlated with the ``elementary-cube monopoles'' {\em in this gauge\/},
and that they signal the presence of physically relevant monopoles.

\section{Silver Pavillion: Maximally Abelian Gauge}

\label{silver}

Let us briefly sketch what lattice simulations using the MAG have
achieved in recent years.

What one does in lattice simulations, is work with the abelian fields
after the gauge fixing only.
So one generates configurations according to the
full SU(2) Haar measure, then one puts these configurations in the chosen gauge,
and subsequently one throws away the non-abelian components $A^{1,2}$:
all observables are constructed using the abelian field $A^3$ only.

For example, one can construct ``abelian Wilson loops'' in this way.
Surprisingly, one finds evidence\cit{Suz89} for the phenomenon of
``abelian dominance'' hypothesized in Ref.\ 6: 
these abelian Wilson loops show confining behaviour with a string tension
compatible with the one obtained from the full SU(2) gauge fields.

Later it was found furthermore that Wilson loops made of the magnetic
monopole fields contained in this abelian component reproduce the string
tension as well (``monopole dominance'')\cit{Suz94}.
Issues related to chiral symmetry breaking have also been studied using
the abelian fields and the monopoles in the MAG\cite{Miy95}.
Let us emphasize that these monopoles are defined using the
elementary-cube definition of Ref.\ 4, 
{\em not\/} by identifying singularities in the gauge fixing procedure.

More recently, studies of monopole effective actions and their renormalization
group behaviour have yielded very interesting results\cit{Suz98}.

Other gauge conditions for the Abelian Projection have not shown convincing
evidence of abelian or monopole dominance (and monopole effective action
results are not available for other gauges than MAG).
The reason is presumably that MAG is a smooth gauge, while the others
are not.
By ``smooth'' we mean, roughly speaking, that (the non-abelian parts of)
the link matrices are close to unity, and do not vary too wildly from one
site to the next.
This kind of smoothness is required for a reliable mapping
$\{ U_{\mu,x}\} \ra \{ A_{\mu,x} \}$
of lattice to continuum gauge fields, which is inherent in the projection
procedure implemented in lattice simulations.
 
The successes of the MAG are overshadowed by an important shortcoming,
however:
{\em It is ambiguous!\/}\@
The gauge condition is implemented by means of a local iterative
minimization algorithm, which often ends up in one of a number of
local minima of the functional (\ref{maglat}).
This means that two gauge equivalent starting configurations may lead
to different gauge fixed configurations.
In other words: gauge invariance is lost, and the physics
extracted from the abelian projection using the MAG becomes unreliable.
These local minima are sometimes called ``Gribov copies''.
However, most of them are ``lattice Gribov copies'',
UV artefacts of the lattice procedure,
which should be distinguished from ``true'' Gribov copies.

Another weak point is that the magnetic monopoles are not detected
by looking for singularities in the gauge fixing procedure in the
spirit of 't~Hooft\cit{DeG80},
but instead by projecting out the abelian component
and constructing monopoles according to the elementary-cube
construction\cit{DeG80} familiar from U(1) theories.
It is not clear that these are always related.
One can consider the physical example of a dyon, to gain some insight
into this relation (see Ref.\ 10 
for a discussion of the dyon
in different abelian projection gauges).  Here one finds\cit{vdS91}
that both definitions point towards a monopole singularity at the origin.
The singularities of the MAG fixing are found to occur when the non-abelian
fields diverge (become of order ${\cal O}(a^{-1})$ on the lattice),
while the cube definition looks for Dirac monopoles: points where the
{\em abelian\/} component blows up.
For the dyon, which is a regular solution, these singularities in the
abelian and non-abelian gauge field components have to coincide, but for 
a general elementary-cube Dirac monopole found in simulations it is not clear
that it comes from a physically relevant monopole.

In fact, one can argue that physical monopoles are suppressed too
strongly by the MAG.  For the (continuum) dyon in MAG (\ref{magcont}), we know
the non-abelian fields $A^{1,2}$ blow up, so the functional (\ref{Scont})
must accomodate a large
contribution from the central region of the dyon.   This goes against its
tendency to minimize this quantity.
So here we have another drawback of the MAG: it unnaturally attempts to
suppress physical monopoles.

The Laplacian method for Abelian Projection will be seen to kill all these
birds with one stone.
It is unambiguous, yields smooth gauge fixed configurations,
does not unduly suppress physical monopoles, and there is evidence that
the elementary-cube monopoles have physical significance.

\subsection{MAG as a spin (glass) model}

\label{spinglass}

To set the scene for the presentation of the Laplacian Abelian Projection in
Sect.~\ref{golden}, let us try to get some insight into what the MAG 
actually does\cit{vdS95,vdS91}.
Going back to the functional $\tilde S$ of Eq.\ (\ref{maglat}),
we note that it can be written as
\bea
\tilde\calS_U(\phi) &=& \sum_{x,\mu}
 \left\{ 1 - \half \Tr [ \Phi_x U_{\mu,x} \Phi_{x+\hat\mu} U^+_{\mu,x}]\right\}
 \label{S2} \\
&=&
 \sum_{x,\mu}
    \{ 1 - \sum_{a,b} \phi^a_x R^{ab}_{\mu,x} \phi^b_{x+\hat\mu} \}
 \, , \label{S3}
\eea
where
\be
R^{ab}_{\mu,x} \ = \ \half \Tr [ \sigma_a U_{\mu,x} \sigma_b U^+_{\mu,x} ]
  \label{Rab}
\ee
is the lattice gauge field matrix in the adjoint representation.
Here $\Phi$ is defined as
\be
\Phi_x \ =\ \Omega^+_x \sigma_3 \Omega_x
       \ =\ \sum_{a=1}^3 \phi^a_x \sigma_a
  \, . \label{Phi}
\ee
It satisfies the constraint
\be
\sum_{a=1}^3 (\phi^a_x)^2 = 1
 \, . \label{phiconstraint}
\ee

$\Phi$ is a very natural variable to parametrize
the gauge transformation $\Omega$:
the original gauge transformation matrix $\Omega_x \in$ SU(2)
is determined up to the residual U(1) gauge invariance only,
but $\Phi_x \in S^2 \simeq SU(2)/U(1)$ is invariant under
$\Omega_x \rightarrow V_x \Omega_x$ 
for diagonal matrices $V_x \in U(1) \ \subset\ SU(2)$:
\be
\Phi_x \ =\ \Omega^+_x \sigma_3 \Omega_x \ \ra\
            \Omega^+_x V^+_x \sigma_3 V_x \Omega_x \ =\ \Phi_x
 \, .\label{Phitrf}
\ee
(In other words, $\Phi$ is neutral with respect to the residual U(1) symmetry.)

Eq.\ (\ref{S3}) is the lattice discretization of the continuum functional
\be
\tilde\calS_A(\phi) \ = \
\int_V \half (D_\mu \phi)^2  \, ,
 \label{glass}
\ee
which could also have been obtained directly from Eq.\ (\ref{Scont}).
This is the covariant kinetic action of a unit-length adjoint spin field in
the background of the gauge field configuration which one wants to
gauge fix.
It is this functional which the MAG attempts to minimize.

Because of the quasi-random nearest-neighbour couplings between the
spins in this action (\ref{S3}), it is similar to a spin glass model.
It is the length-1 constraint on the spin variables which is responsible
for the difficulties in solving this minimization problem.

\section{Golden Pavillion: Laplacian Abelian Projection}

\label{golden}

The Laplacian method\cit{vdS97}%
\protect\footnote{This follows a similar idea in the context of
Landau gauge fixing on the lattice\protect\cit{Vin92}.}
is based on relaxing the length-1 constraint
on the $\phi$ fields in the minimization problem
(\ref{S3},\ref{glass}) of the MAG.

In this case the ``spin glass action'' (\ref{S3},\ref{glass}) becomes an
unconstrained quadratic form which is minimized by the
the lowest-lying eigenvector
of the covariant Laplacian operator $-\Box(R) = - \sum_\mu D_\mu(R) D_\mu(R)$.
(We do of course maintain an overall normalization on the
eigenvectors.)\@
On the lattice,
$-\Box(R)$ is the $3V\times 3V$ real symmetric matrix
\be
-\Box^{ab}_{xy}(R) \ = \
  \sum_\mu \left( 2 \delta_{xy} \delta^{ab}
  \ - \ R_\mu^{ab}(x) \delta_{y,x+\hat\mu}
  \ - \ R_\mu^{ba}(x) \delta_{y,x-\hat\mu}
      \right)
 \, .
 \label{Boxlat}
\ee
Its eigenvalues and eigenvectors can be computed numerically in an
unambiguous way and to arbitrary precision.
One can use the Lanczos method (which in addition to eigenvalues
can provide the corresponding eigenvectors), or a Rayleigh-Ritz method,
for example.

The only situation in which an ambiguity might arise is when
the lowest eigenvalue is degenerate.
The two (or more) corresponding eigenvectors
would lead to different (but all smooth) gauge field configurations,
both satisfying the gauge condition exactly.
One would have a true, physical Gribov copy (as opposed to the
``lattice Gribov copies'' afflicting the MAG).
However, this set of true Gribov copies has measure zero, and they
are never encountered in practice:
The two lowest eigenvalues are always found to be separated by an
amount which is orders of magnitude larger than the numerical precision
with which they can be resolved.

It remains to be discussed how the lowest mode of the Laplacian determines
the gauge fixing.
By relaxing the length-1 constraints on the fields $\vec\phi_x$ we have of
course abandoned their straightforward interpretation in terms of
gauge transformation matrices $\Omega_x$ through Eq.\ (\ref{Phi}).

There are two equivalent ways to see how the lowest eigenvector $\phi^a_x$
defines the required gauge transformation $\Omega_x$.
One way is to normalize $\phi^a_x$ site by site, by writing
\bea
\phi^a_x &=& \rho_x \hat\phi^a_x
 \, , \label{phinormal} \\
\rho_x &=& \left( \sum_{a=1}^3 (\phi^a_x)^2 \right)^{1/2}
 \, , \label{rhonormal}
\eea
and to use the normalized
vector to obtain $\Omega_x$ (up to its U(1) arbitrariness of course):
\be
\Omega^+_x \sigma_3 \Omega_x \ = \ \sum_{a=1}^3 \hat\phi^a_x \sigma_a
 \, .  \label{Omeganormal}
\ee
The other way is to realize that the matrix which rotates the
3-vector $\vec\phi_x$
into the abelian direction (which we chose to be the 3-direction),
is nothing but the required gauge transformation $\Omega_x$ in the
adjoint representation.
In this picture, the U(1) arbitrariness
corresponds to rotations about the direction of $\vec\phi_x$.

This procedure is ill-defined only when $\rho_x =0$ for some $x$.
At such points there is no way to decide which $\Omega_x$ to take.
In other words, since $|\vec\phi_x| = 0$, the symmetry is not broken
to U(1) but one retains the full SU(2).
This is a singularity in the gauge fixing procedure,
of the type that identifies a magnetic monopole\cit{tHo81}.
Note that this is very similar to the zero of the Higgs field in the
true 't~Hooft-Polyakov monopole in the Georgi-Glashow model!
This is a promising feature, suggesting  a direct relation of these
gauge fixing singularities with physical monopoles.
We shall discuss this in detail in Sect.\ \ref{rhosubs}.

\subsection{Laplacian Abelian dominance}

\label{abdom}

We will present preliminary results concerning abelian dominance using the
Laplacian Abelian Projection.
\begin{figure}[b]
\vspace*{5mm}

\centerline{
\epsfxsize=0.5\textwidth
\epsfbox{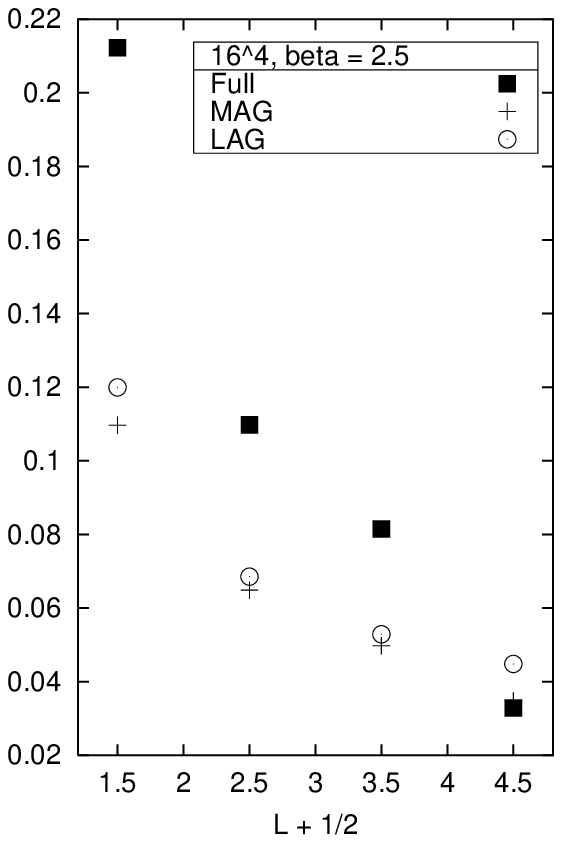}
\epsfxsize=0.5\textwidth
\epsfbox{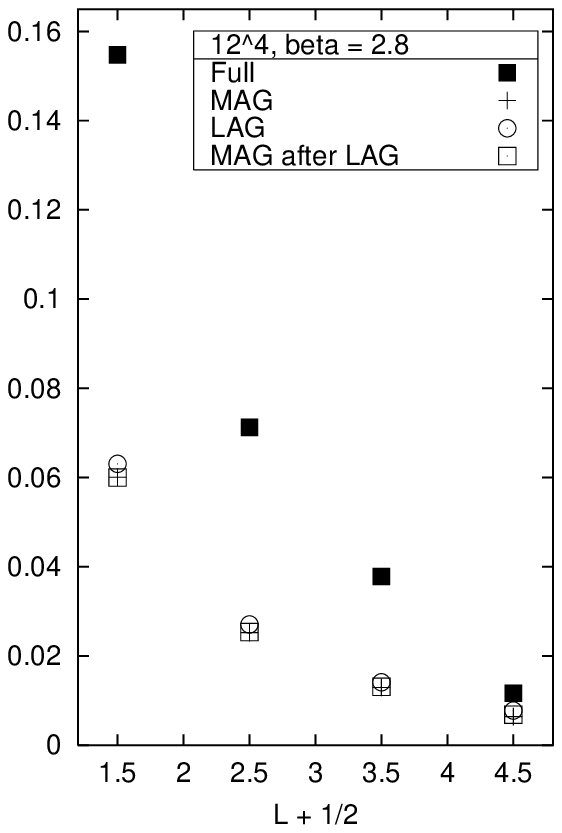}
}
\caption{Abelian dominance: comparison of full and abelian ``square''
Creutz ratios for two SU(2) simulations.  The Laplacian method (LAG) seems
to be slightly ``more abelian dominant'' than the Maximally Abelian Gauge
(MAG).}
\label{fig1}
\end{figure}

Fig.\ \ref{fig1} shows the ``square'' Creutz ratios
\be
\chi(L+1/2,L+1/2) \ = \ - \ln \frac{\langle W(L,L) \rangle
                             \langle W(L+1,L+1) \rangle}
                            {\langle W(L,L+1) \rangle
                             \langle W(L+1,L) \rangle}
 \label{creutz}
\ee
($W(R,T)$ denotes an $R\times T$ Wilson loop)
plotted against $L+1/2$, for two SU(2) simulations.
Shown are the full SU(2) Creutz ratio and the ``abelian Creutz ratios'',
formed from the abelian gauge field components after abelian projection
using the MAG and the Laplacian Abelian Gauge (LAG).
The errors (not shown) on the data points at $L+1/2 = 4.5$ are large,
especially for the $16^4$ simulation at $\beta=2.5$.

One observes that $\chi_{\rm LAG}$ is about the same as $\chi_{\rm MAG}$.
Both $\chi_{\rm LAG}$ and $\chi_{\rm MAG}$ approach $\chi_{\rm Full}$ at
large distance, but they are smaller than $\chi_{\rm Full}$ at short distances.
This is in agreement with the fact that the smaller Coulomb term in the
abelian potential leads to a smaller slope $dV/dr$ in that regime.
Thus we find that the LAG shows abelian dominance just like the MAG.
In fact, the data seem to suggest that $\chi_{\rm LAG}$ may even lie
slightly closer to $\chi_{\rm Full}$ than $\chi_{\rm MAG}$.
However, more accurate calculations are needed before this can be
concluded with confidence.

The $12^4$ graph shows in addition the abelian Creutz ratios for gauge
fixing to the MAG after preconditioning the configurations with the LAG.
They seem to be roughly equal to $\chi_{\rm MAG}$ as calculated without
LAG-preconditioning, and again slightly smaller than $\chi_{\rm LAG}$.
(This gives some indication of the size of ambiguities in MAG gauge fixing.)\@

Hence we find first evidence that the LAG method passes the test of
abelian dominance; more detailed computations as well as a study of
monopole dominance are under way.

\subsection{The Higgs-like degree of freedom $\rho$}

\label{rhosubs}

We want to examine in some more detail
the origin and physical relevance of the Higgs-like field $\phi^a_x$,
and in particular its length $\rho_x$,
which emerges in the Laplacian gauge fixing procedure.

First of all, note that it is a gauge-invariant property of the configuration,
in the sense that (unlike the lowest eigenvector $\phi^a_x$ itself)
it does not depend on the initial gauge of the full SU(2) configuration.

We found that zeroes in the field $\phi$ correspond to singular points in
the gauge fixing, and hence to monopoles in 't~Hooft's Abelian Projection
picture.
This is reminiscent of the Georgi-Glashow model, where 3-dimensional
vortex-like configurations of the Higgs field (with zeroes in their centre)
are characteristic of the presence of a 't~Hooft-Polyakov (tHP)
magnetic monopole.
One wonders how far this analogy can be extended to the Higgs-like field
that emerges in the Laplacian gauge (LAG).

Let us consider what form a classical dyon takes in the LAG.
The answer can be obtained by the following argument.
The LAG attempts to minimize $\tilde{\cal S}_A(\phi) = \int(D_\mu\phi)^2$ in
the background of the gauge field configuration of the classical dyon.
The spatial gauge field components of this dyon equal those of the tHP
monopole in the Georgi-Glashow model, while their 0$^{\rm th}$ or 4$^{\rm th}$
components differ.
If we forget this latter difference for a moment, we conclude that
since $\tilde{\cal S}_A(\phi)$ has exactly the form of the Higgs kinetic
action in the Georgi-Glashow model, the static Higgs field part of the
tHP monopole solution in the BPS limit is exactly what the Laplacian
eigenvector $\phi_x$ should look like for the dyon in the LAG.
Finally, it is clear that the non-zero 4$^{\rm th}$
component of the dyon solution does
not change this conclusion: it (also) has the form of the static
Higgs field of the tHP monopole in the BPS limit, so its contribution
to $(D_4\phi)^2$ is zero.

So the lowest eigenvector of the covariant Laplacian in the background
of a classical dyon of scale $\mu$ in radial gauge is given by
\bea
\phi^a_x &=& \delta_{ak} \hat x_k \rho(r) \, , 
 \label{dyonphi} \\
 \rho(r) &=& \mu \frac{\cosh \mu r}{\sinh \mu r} - \frac{1}{r} \, .
 \label{dyonrho}
\eea
This implies first of all that the Laplacian gauge is reached by rotating
the Higgs field into the 3-direction.  This is simply unitary gauge, and
the 3$^{\rm rd}$ component of the gauge field configuration will have the
form of a Dirac monopole field.
(So the elementary-cube procedure for detecting monopoles will identify  
it, just like it does in the MAG.)\@
At the same time, we see that, in Laplacian gauge, the classical dyon
shows up as a region (of extent $\mu^{-1}$) of small $\rho$, with
a zero in its centre.

This also allows us to deduce that the dyon is not suppressed in an
unnaturally strong fashion in the LAG,
as opposed to the MAG case (cf.\ the discussion
in Sect.\ \ref{silver}).
The point is that the quantity that is minimized in the LAG is not
$\tilde\calS_U(\Omega)$ of Eq.\ (\ref{maglat}) but
\be
\tilde\calS_U(\Omega,\rho) \ = \
2 \sum_{x,\mu} \rho_x\,\left[ (u^1_{x,\mu})^2 + (u^2_{x,\mu})^2) \right]
  \,\rho_{x+\hat\mu} 
 \label{maglatrho}
\ee
(with an overall normalization imposed on $\rho$).
Here we have written the link matrices $U_{\mu,x}$ as
$u^0_{x,\mu} + i u^a_{x,\mu} \sigma^a$.
In the continuum formulation, this corresponds to
\be
\tilde\calS_A(\Omega,\rho) \ = \ \frac12 \int_V \rho^2 \,
   \left[ \left(A_\mu^{(\Omega)\,1}\right)^2
 \ + \ \left(A_\mu^{(\Omega)\,2}\right)^2 \right]
 \, . \label{Scontlag}
\ee
The value of $\rho$ tends to zero as the non-abelian field components
$A^{1,2}$ blow up when the singularity is approached, thus moderating
the contribution of the dyon to the functional (\ref{Scontlag}).
So the LAG automatically arranges things such that dyons are
accomodated in the Laplacian gauge without a large penalty.

As a matter of fact, this may also explain the somewhat higher monopole
density found using the LAG\cit{vdS97}: physical monopoles (dyons) are not
too vehemently suppressed, unlike what is the case for the MAG.

This leads us to the following question:
what is the physical significance of the
magnetic monopoles found using the conventional
elementary-cube procedure\cite{DeG80},
in LAG and MAG?

For the LAG, it is natural to ask, in view of the above,
whether regions of small $\rho$ (implying the presence of physical monopoles,
identified as gauge fixing singularities in the Laplacian Abelian Projection)
are correlated with the location of monopoles defined by means of
the elementary-cube procedure.
This correlation was visualized in the video shown at the workshop,
and included in the CD-ROM version of these Proceedings\cit{webvideo}.
For the MAG, we studied a similar correlation.

\begin{figure}[p]
%
%
\vspace*{17cm}
\caption{Correlation between small-$\rho$ regions and elementary-cube
monopoles for Laplacian Abelian Projection.
The two panels are different visualizations of the same situation, taken
from a $12^3\times 4$ simulation at $\beta=3.0$, with periodic boundary
conditions.
The box shows the $12\times 12\times 4$ $x$-$y$-$t$ hyperplane at some
(dual) value of $z$.
The blue regions correspond to small $\rho$ (in the bottom panel, only
some of the boundaries are shown).
The orange tubes represent magnetic monopole currents running in the 
$x$-$y$-$t$ hyperplane,
the small yellow and red balls are
monopole currents in the $z$-direction orthogonal to this hyperplane.
}
\label{videoframe}
\end{figure}

Fig.\ \ref{videoframe} shows one frame from this video.
The two panels correspond to one and the same situation, visualized
in different ways to highlight different features.

The main observations are:

$i$)\ \ 
The regions of small $\rho$ (smaller than a cutoff of 0.4, relative to
an average of 1.0) form large-scale structures, rather than looking like
widely scattered UV noise.

$ii$)\ \ 
There is a clear correlation between these regions and the location of
monopoles identified using the elementary-cube method.
In Fig.~\ref{videoframe} this is particularly clear in the bottom panel,
where one sees that the monopoles are enclosed in the small-$\rho$
regions.

$iii$)\ \ 
Furthermore, although somewhat unrelated to the present discussion, it is
observed that (in the high-temperature phase) both small-$\rho$ regions
and monopole loops appear to align along the euclidean time direction.
This is in agreement with the prediction of Ref.\ 3, 
which describes the QCD vacuum as a condensate of dyon-like configurations.
Should this feature survive in full QCD (cf.\ the conclusions
of Ref.\ 15), 
then it would be
at odds with expectations based on instanton liquid models\cit{Shu97}
which predict instanton-antiinstanton molecules (dipoles) aligned along
the euclidean time direction in the full theory.

$iv$)\ \ 
For comparison, we looked for a similar correlation in the case of the MAG.
Since we do not have a quantity like $\rho$ at our disposal in this gauge,
we considered instead the correlation of the monopole locations with
the ``non-abelian residue'', {\em i.e.}, the local value of the 
integrand of (\ref{maglat}).
The behaviour is strikingly different.
In contrast with what is observed in the LAG,
this non-abelian residue looks like purely short-range
UV noise (countless small blue ``splinters'' as opposed to the few large blue
blobs in Fig.~\ref{videoframe}),
without any long-distance structure.  No correlation with
the monopole locations is observed either.

These observations lead to the conclusion that the $\rho$ field signals
important physical structure, and that the monopoles found in the
LAG have physical meaning.  No such evidence is available for the MAG.

\section{Meditation}

We have presented a new, ``Laplacian'' method (LAG) to perform the
Abelian Projection on the lattice, which has several advantages
over the popular Maximally Abelian Gauge (MAG).

Most importantly, it is not afflicted with the ambiguities inherent to
the local iterative minimization procedure needed in the MAG.
It is a manifestly unambiguous procedure, which leads to gauge fixed
configurations endowed with a natural (Laplacian) kind of smoothness.

In addition, it treats physical monopoles mildly, contrary to the MAG.
The large penalty which they receive in the MAG and which gives rise to
a strong suppression, is compensated by the Higgs-like field in the LAG,
which automatically arranges itself so as to accomodate
the monopoles comfortably.
We conclude that the MAG suppresses physical monopoles {\em too\/} strongly.
Or in other words, it ignores important physical information which is
contained in the variable $\rho$.

Finally, there is evidence that, {\em in the LAG,} the monopoles
found using the elementary-cube method are
correlated to those defined by the singularities in the partial gauge fixing.
This evidence was obtained from an analysis of the classical dyon on the
one hand, and from studying the correlations between the Higgs-like
variable $\rho$ and magnetic monopoles in simulations on the other hand.

Furthermore, we found first evidence for abelian dominance using the Laplacian
Abelian Projection, similar to the MAG.
This study will be pursued in more depth and extended to monopole dominance
in the near future.

\section*{Acknowledgements}

It is a pleasure to thank the organizers of YKIS\,'97 for an excellently
organized and very enjoyable workshop.
D\=omo arigat\=o gozaimashita!
I also want to express my thanks to Ph.~de Forcrand for
stimulating advice and discussions,
and to R.~Peikert for his expertise in producing the data visualization video
presented at the workshop and included in the CD-ROM version of these
Proceedings.


\end{document}